\RequirePackage{lineno}
\documentclass{article}
\usepackage{arxiv}
\usepackage{cleveref}
\usepackage{graphicx}
\usepackage{amsmath,amssymb}
\usepackage{siunitx}
\usepackage{booktabs}
\usepackage{multirow}
\usepackage{comment}
\usepackage{hyperref}

\title{Predicting the Neutrino Mass Ordering \\ Using Neural Networks}

\usepackage{authblk}
\author[1]{
   \href{https://orcid.org/0000-0002-0424-7903}{T.J.C. Bezerra}
  }
\author[1]{
   \href{https://orcid.org/0000-0001-8035-7162}{L. Asquith}
  }
 \author[1]{
   \href{https://orcid.org/0000-0002-7535-5856}{E. Bannister}
  } 
  \author[1$\dagger$]{
   \href{https://orcid.org/0000-0002-7221-1910}{W. Shorrock}
  }  

\affil[1]{Department of Physics and Astronomy, University of Sussex, Brighton, BN1 9QH, United Kingdom}

\affil[$\dagger$]{Now at Universal Quantum Ltd., Haywards Heath, RH16 1XQ, United Kingdom}

\begin{document}
\nolinenumbers

\maketitle

\begin{abstract}

Determining the neutrino mass ordering remains a central open problem in particle physics. While next-generation long-baseline experiments are expected to resolve this question, current data provide limited sensitivity because the spectral differences between normal and inverted ordering are subtle and entangled with parameter degeneracies. We investigate a machine-learning strategy for mass-ordering determination using a feed-forward neural-network classifier trained on synthetic long-baseline datasets generated with three-flavour oscillation probabilities, matter effects, and statistical fluctuations. We evaluate the classifier against standard $\chi^2$ and $\log\mathcal{L}$ approaches using common discrimination metrics, including receiver-operating-characteristic curves, to quantify sensitivity and to illustrate how operating points can be selected to prioritise purity or efficiency. We find that the neural network achieves performance comparable to conventional fits for the scenarios studied, providing a flexible, independent cross-check of established analyses. The framework can be extended to incorporate systematic uncertainties and to explore joint inference of oscillation parameters, and it may also serve as a pedagogical tool for introducing machine-learning methods in neutrino physics.
\end{abstract}

\section{Introduction}
\label{sec:intro}

Neutrino oscillations provide compelling evidence for physics beyond the Standard Model of particle physics, revealing that neutrinos have mass and mix across flavours~\cite{suekane:2015}. The oscillation framework describes the neutrino mass eigenstates ($\nu_1$, $\nu_2$, $\nu_3$), which govern propagation, as quantum superpositions of the flavour eigenstates ($\nu_e$, $\nu_\mu$, $\nu_\tau$), which govern interactions. In this framework, a neutrino produced with flavour $\alpha$ can be detected as a different flavour $\beta$ with a non-zero probability, $P(\nu_\alpha \rightarrow \nu_\beta) > 0$, where $\alpha,\beta \in \{e,\mu,\tau\}$. This probability depends on neutrino energy and propagation distance, oscillating sinusoidally between 0 and 1, hence the term ``neutrino oscillations''. The standard description of this phenomenon involves six independent parameters: three mixing angles ($\theta_{13},~\theta_{12},~\theta_{23}$) related to the oscillation amplitudes, two independent mass-squared differences ($\Delta m^{2}_{21}$ and $\Delta m^{2}_{32}$)\footnote{ $\Delta m^{2}_{21}$ and $\Delta m^{2}_{32}$ are also known as the "solar splitting", $\Delta m^2_{\odot}$, and "atmospheric splitting", $\Delta m^2_{\rm{atm}}$, respectively.} related to the oscillation frequency, and one charge-parity violation phase ($\delta_{CP}$), which describes whether neutrinos and antineutrinos oscillate differently. 

One of the remaining open questions in neutrino physics is the ordering of the neutrino mass states, known as the neutrino mass ordering\footnote{Some authors use the term `mass hierarchy'.} (MO)~\cite{vogel:2015}. Although solar neutrino experiments have established the ordering of the first two states as $m_1 < m_2$~\cite{borexino:2018}, it remains unknown whether the third mass state is heavier (normal ordering, NO) or lighter (inverted ordering, IO) than the other two. Determining the neutrino mass ordering is central to advancing the understanding of fundamental physics. The ordering influences a wide range of phenomena, from the interpretation of neutrinoless double beta decay experiments~\cite{nu02b:2016}, which are critical for probing whether neutrinos are Majorana particles, to the modelling of supernova neutrino flavour conversions and nucleosynthesis processes~\cite{supernova:2014}. In cosmology, the mass ordering constrains the sum of neutrino masses, which affects large-scale structure formation and cosmic microwave background measurements~\cite{cosmology:2012}. These implications underscore why determining the neutrino mass ordering is a priority for next-generation neutrino experiments and complementary approaches, including machine-learning techniques.

Several strategies have been developed to determine the neutrino mass ordering, each exploiting distinct physical effects. One approach relies on neutrino propagation through matter, where interactions between neutrinos and electrons in their path introduce an additional phase that modifies oscillation probabilities. This phenomenon, known as the Mikheyev--Smirnov--Wolfenstein (MSW) effect~\cite{wolfenstein:1978,mikheyev:1985}, enhances sensitivity to the mass ordering in long-baseline accelerator experiments and atmospheric neutrino measurements. A second approach leverages the subtle difference between the two larger mass-squared splittings, $\Delta m^2_{31}$ and $\Delta m^2_{32}$ under the different MO assumptions, which affects oscillation patterns at specific energies and baselines~\cite{petcov:2002}. Both methods require carefully designed experiments with precise control of systematics and high statistics, as the signatures of the ordering are inherently small. Beyond these direct approaches, complementary information can also be obtained from cosmological observations and neutrinoless double beta decay searches, which constrain the absolute neutrino mass scale and provide indirect sensitivity to the ordering.

Among these strategies, this work focuses on the matter-effect approach implemented in long-baseline accelerator experiments. In these facilities, protons are accelerated to $\mathcal{O}$(GeV) energies and collided with a stationary target, producing secondary mesons such as pions and kaons. Magnetic horns are then used to focus these particles and select the desired charge before they decay into neutrinos within a dedicated decay pipe. This process results in a beam predominantly composed of $\nu_{\mu}$ or $\bar{\nu}_{\mu}$, which is monitored by dedicated detectors, often called near detectors. By placing a detector hundreds of kilometres away, it becomes possible to measure how many of these muon neutrinos remain and how many have oscillated into electron neutrinos\footnote{or $\tau$ neutrinos; however, the $\tau$ charged-current production threshold is $\sim$3.5 GeV, above typical long-baseline beam energies.}. The oscillation parameters are extracted by comparing the observed energies with the expected non-oscillated energy spectra. The first generation of experiments using this approach included K2K and MINOS in the early 2000s~\cite{k2k:2006,minos:2013}. Currently, the second-generation experiments, NOvA and T2K, are completing data collection~\cite{nova_T2K:2025}, while the third-generation facilities, DUNE and Hyper-K, are expected to begin operation in the late 2020s~\cite{dune:2015,hyperk:2018}.

Traditional approaches to MO determination rely on statistical fits to oscillated energy spectra, typically using $\chi^{2}$ or log-likelihood minimisation techniques. While robust, these methods often exhibit limited discrimination power due to the subtle differences between NO and IO scenarios, as observed in experiments such as T2K, where the spectral variations are small. In the case of NOvA, the ambiguity in MO determination is compounded by degeneracies with the CP-violating phase, making it difficult to draw firm conclusions from current data, even in a combined analysis~\cite{nova_T2K:2025}. These challenges motivate the exploration of alternative approaches, including machine-learning techniques, which have the potential to test for, or identify, complex correlations in oscillation energy spectra that may be difficult to capture with traditional statistical fits.

In this work, we investigate a machine-learning approach based on a neural network (NN) classifier trained on synthetic datasets that replicate published experimental conditions. The training samples are generated using a fast calculator of three-flavour neutrino oscillation probabilities, including the MSW matter effect, and incorporate statistical fluctuations. The goal is to assess whether machine learning can improve sensitivity to the neutrino mass ordering compared to standard $\chi^{2}$ or likelihood-based methods. Our model is evaluated against these traditional techniques to quantify potential gains in discrimination power.

The remainder of this paper is organised as follows. \Cref{sec:methods} describes the methodology used to generate synthetic datasets and the NN architecture employed for classification. It also presents the training procedure and evaluation metrics. \Cref{sec:results} presents the results, focusing on classification performance and ROC curve analysis. This section also includes comparisons with standard log-likelihood fitting techniques. Finally, \Cref{sec:concl} summarises our findings and suggests directions for future work, highlighting the potential of machine learning to complement traditional analysis in neutrino physics.

\section{Simulated Datasets \& Methods}
\label{sec:methods}

This study uses the NOvA (\textbf{N}uMI \textbf{O}ff-axis $\boldsymbol{\nu}_e$ \textbf{A}ppearance) experiment~\cite{nova:2007} as a reference framework for generating synthetic neutrino energy spectra and calculating oscillation probabilities. The NOvA Near Detector (ND) and Far Detector (FD) are separated by a distance of 810~km, with the ND located at the Fermilab accelerator complex where (anti)neutrinos are produced. Beams of both neutrinos and antineutrinos are used in the NOvA experiment, hereafter collectively referred to as ``(anti)neutrinos'' unless stated otherwise. The detectors were constructed using PVC extrusions to contain cells of liquid scintillator~\cite{Talaga:2016rlq}. Each cell contains a wavelength-shifting optical fibre that transports scintillation light to photosensors, enabling the measurement of light intensity and timing for each data-acquisition trigger. The cells are arranged to provide two orthogonal 2D projections of neutrino interactions: top-bottom and east-west. These measurements form an event ``image'' that allows reconstruction of the interaction type (charged-current, CC, producing an electron or muon, or neutral-current, NC, with no charged lepton), as well as the neutrino flavour and energy.

\begin{figure}[htpb!]
    \centering
    \includegraphics[width=0.65\textwidth]{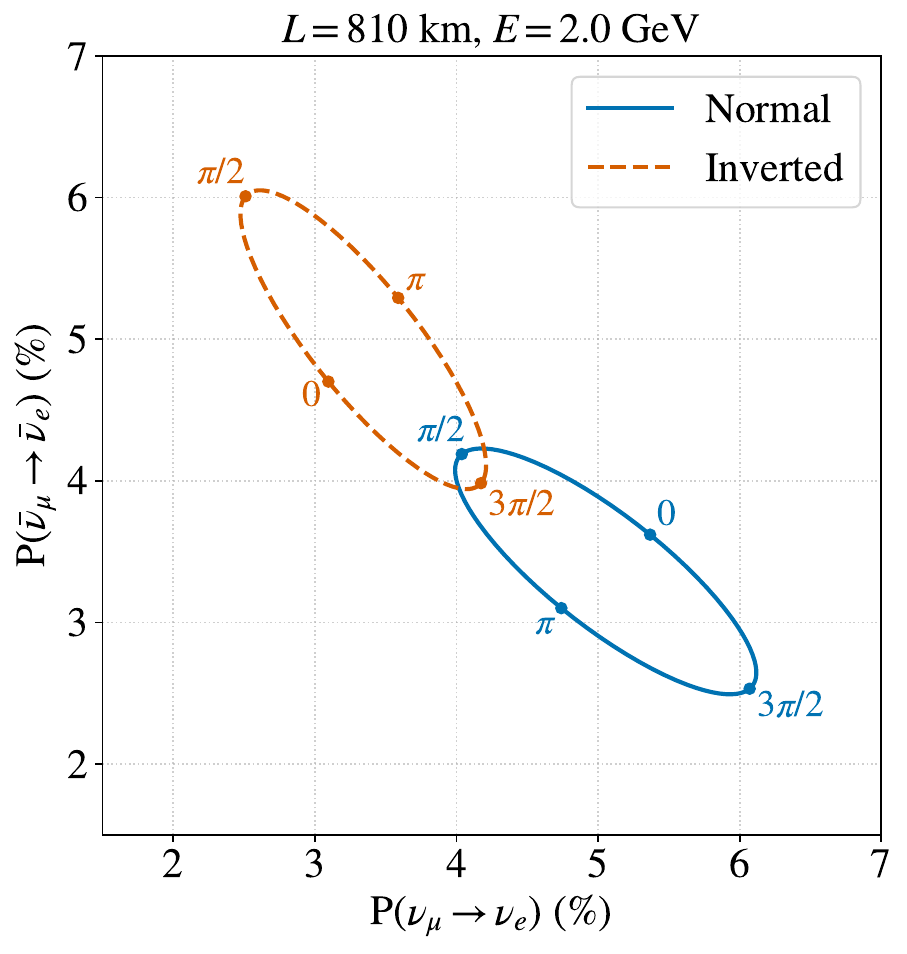}
    \caption{The probability of muon-to-electron neutrino oscillations for neutrinos versus antineutrinos with the NOvA baseline of L = \SI{810}{\km} and a fixed typical source neutrino energy of \SI{2}{\GeV}. The blue (dashed, orange) ellipse is for normal (inverted) mass ordering, with the corresponding oscillation parameters fixed to the values in \Cref{tab:params}. The points along the ellipses are for different values of $\delta_{\text{CP}}$ in the $[0,2\pi]$ range and the markers indicate a few key values of $\delta_{\text{CP}}$.}
    \label{fig:nu_vs_anu_prob}
\end{figure}

The motivation for using NOvA lies in its sensitivity to matter effects and the degeneracies between mass ordering and the CP-violating phase. 
For example, \Cref{fig:nu_vs_anu_prob} illustrates the electron neutrino and antineutrino appearance probabilities for an electron (anti)neutrino from a muon (anti)neutrino with an energy of \SI{2}{\GeV} travelling \SI{810}{\km} from the source. Each ellipse corresponds to a choice of mass ordering, with its shape determined by varying $\delta_{CP}$ from 0 to $2\pi$. As shown, certain probability values can be obtained under different assumptions of MO and $\delta_{CP}$, creating degeneracies that complicate interpretation. The partial separation between the ellipses arises from matter effects; in vacuum, the ellipses would overlap almost completely. The parameter values used in \Cref{fig:nu_vs_anu_prob} are from the PDG~\cite{ParticleDataGroup:2024cfk}, as given in \Cref{tab:params}.

\renewcommand{\arraystretch}{1.5}

\begin{table}
\caption{Summary of the oscillation parameters used in this study. The middle column (\textbf{Value Used}) lists the central values or ranges used in  \Cref{subsec:MO_estimate}. The right column (\textbf{PDG Value}) lists the combined best-fit values and uncertainties from many experiments, provided by the Particle Data Group~\cite{ParticleDataGroup:2024cfk}. In most cases, the approximate PDG central values are used in this study; however, the NO $\delta_{CP}=0.5$ was chosen to maximise the ambiguity between MO scenarios (\Cref{fig:nu_vs_anu_prob}).}
\centering
\begin{tabular}{
l l r r
}
\toprule
\multicolumn{2}{c}{\textbf{Parameter}} &
\textbf{Value Used} &
\textbf{PDG Value} \\[1ex]
\midrule

\multirow{2}{*}{$\delta_{CP}$ ($\times \pi\, \si{\radian}$)}  & 
NO
&
$0.5$  & 
$1.21^{+0.19}_{-0.22}$\\
& 
IO
&
$1.5$  & 
$1.58^{+0.15}_{-0.16}$\\

$\sin^2\theta_{12}$  &
& 
$0.307$ & 
 $0.307 \pm 0.012$\\[1ex]
\hline

$\sin^2\theta_{13}$  &
& 
$0.0216$ & 
 $0.0216\pm 0.0006$\\[1ex]
\hline

\multirow{2}{*}{$\sin^2\theta_{23}$}&
 NO & 
  $0.535$ &
$0.534^{+0.015}_{-0.019}$\\ 

& IO & 
  $0.535$ &
$0.537\pm 0.020$ \\[1ex]
\hline

$\Delta m^2_{21}$ ($\times 10^{-5}$ eV$^2$)&& 
$7.53$ &
$7.50\pm 0.19$\\ [1ex]
\hline
\multirow{2}{*}{
$\Delta m^2_{32}$ ($\times 10^{-3}$ eV$^2$)}& NO & 
$[2.33,2.57]$ &
$2.451\pm 0.026$  \\ [1ex]
& IO & 
$[-2.65,-2.41]$&
$-2.527\pm 0.034$ \\ 

 \midrule
\bottomrule
\end{tabular}

\label{tab:params}
\end{table}

\subsection{Data Generation}
\label{subsec:data_gen}

For the simulations in this study, the non-oscillated (anti)neutrino energy spectra are approximated by Gaussian distributions with mean $\mu = 1.95$~GeV and standard deviation $\sigma = 0.35$~GeV, normalised to correspond to ten years of data-taking at NOvA’s far detector~\cite{nova:2025}. Approximating the unoscillated spectrum by a Gaussian provides a simple toy model that retains the overall peak location and spread of the NOvA-like energy distribution, allowing us to focus on ordering-dependent spectral distortions rather than flux-modelling details. 

Oscillation probabilities are computed using NOvA’s baseline of 810~km, incorporating matter effects with an assumed constant Earth density of 2.84~g/cm$^{3}$~\cite{nova:2022}. These choices ensure that the synthetic datasets closely resemble realistic experimental conditions and enable meaningful comparisons with published results.

These energy spectra are intended as a controlled toy model. We do not include detector response effects (energy smearing, efficiencies), backgrounds, or other systematic uncertainties. The goal is to isolate the information content of binned energy spectra under statistical fluctuations and matter effects.

The oscillated energy distributions are created by computing the oscillation probabilities for $\nu_\mu$, $\bar{\nu}_\mu$, $\nu_e$, and $\bar{\nu}_e$ using the \texttt{NuFast} package~\cite{nufast:2024}, which provides fast three-flavour oscillation calculations in matter. To obtain bin-averaged event predictions, the binned non-oscillated energy distributions are multiplied by the \texttt{NuFast} probability within each energy bin, integrating numerically to ensure that oscillation effects are accurately captured across the bin width.

For the primary goal of comparing the LLH and NN techniques for estimating the MO, oscillated distributions are generated for both MO scenarios and for a range of \dmatm\ values, with the other parameters of the oscillation probability fixed to the values given in \Cref{tab:params}. The \dmatm\ values are drawn from a uniform distribution centred on the PDG value for the relevant MO and with a range of $\pm 5\%$, chosen to provide full coverage of the current precision on this parameter, which is $\approx 1.5\%$~\cite{nova:2025}.  
A separate set of energy spectra is produced for a benchmark study estimating the two-flavour mass-squared splitting; these are described in \Cref{subsec:benchmark}.

\begin{figure}[htpb!]
    \centering
    \includegraphics[width=0.8\textwidth]{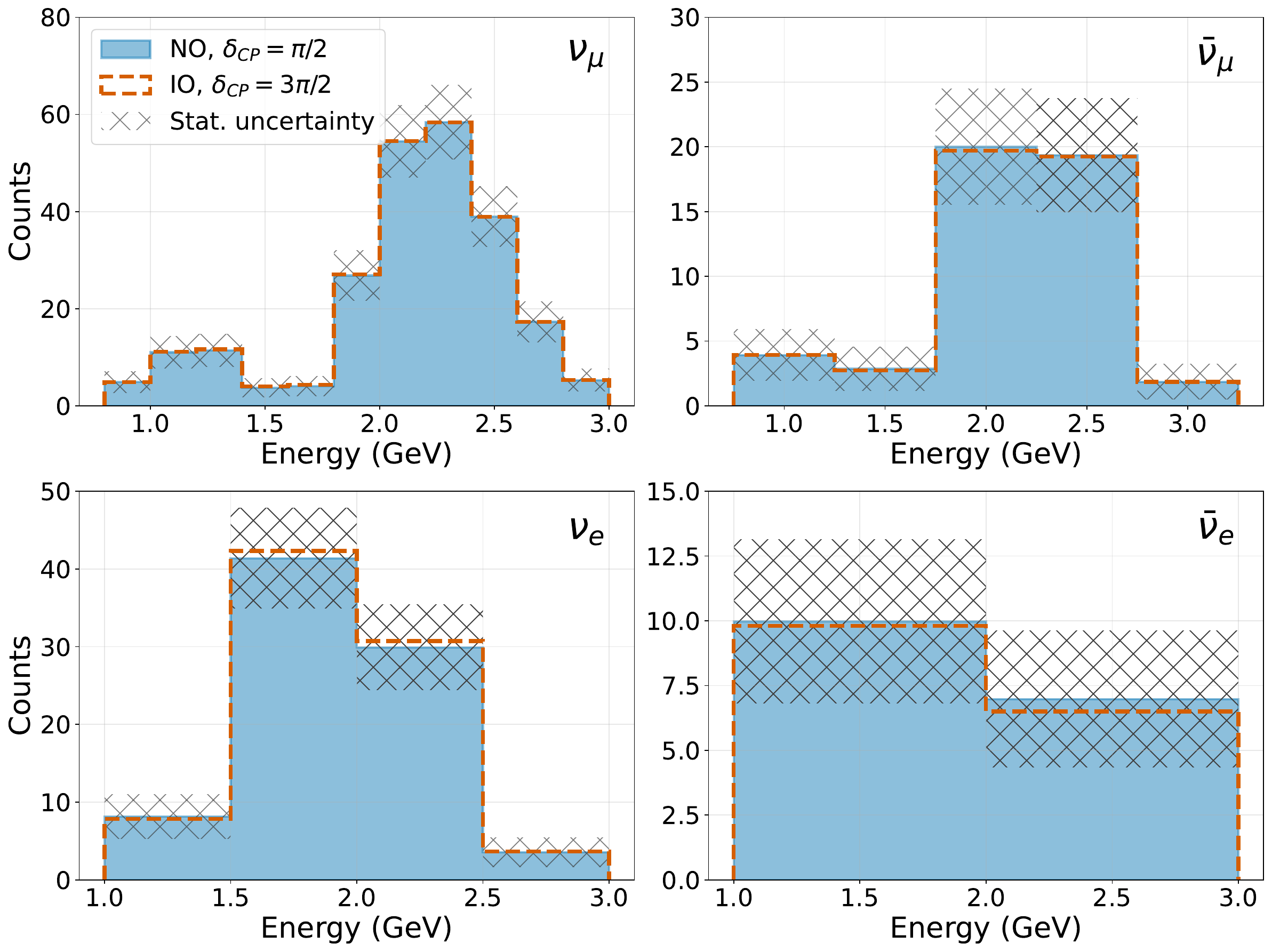}
    \caption{Oscillated energy spectra of $\nu_\mu$, $\bar{\nu}_\mu$, $\nu_e$, and $\bar{\nu}_e$ generated using the parameter values in \Cref{tab:params}. The blue filled histograms (orange dashed-outline) represent normal (inverted) ordering. The hatched boxes indicate statistical uncertainties.}
    \label{fig:hists}
\end{figure}

A given MO and \dmatm\ has four oscillated energy histograms, one for each of the $\nu_\mu$, $\bar{\nu}_\mu$, $\nu_e$, and $\bar{\nu}_e$ energy distributions. \Cref{fig:hists} shows an example set for IO and NO scenarios for the PDG values of $\Delta m^2_{32}$ in \Cref{tab:params}. The differences between the mass-ordering hypotheses are small and lie within the statistical uncertainty band, illustrating the difficulty faced by a binned likelihood fit in discriminating models. The NN, unlike the likelihood fit, has the potential to exploit the behaviour of the models under statistical fluctuations across bins in the entire set of histograms. The binning is chosen for simplicity and to ensure more than two counts per bin, resulting in 22 bins in total for the combined set of four histograms. The array of 22 bin counts then has Poisson fluctuations applied. One million samples of Poisson fluctuated arrays of bin counts are generated for each MO.

\subsection{Parameter Estimation Techniques \& Evaluation Metrics}

The primary goal of this study is to compare the MO estimation of a trained NN to that of a standard LLH fit, based on the simulated data samples described above. 

 The output is either a prediction of the two-flavour mass-squared splitting $\Delta m^2$ (\Cref{subsec:benchmark}) or the three-flavour MO label (\Cref{subsec:MO_estimate}). 
 The mass ordering label is encoded as 1 for NO and 0 for IO, framing the task as a binary classification problem. The predictions from the NN are compared with those obtained using a standard log-likelihood (LLH) fit, implemented as a one-dimensional bounded Brent search via \texttt{scipy.optimize.minimize\_scalar} (\texttt{method="bounded"}) with bounds $(1,\,5) \times 10^{-3}\,\mathrm{eV}^2$ for the NO hypothesis and $(-5,\,-1) \times 10^{-3}\,\mathrm{eV}^2$ for the IO hypothesis, applied to the same input histograms, enabling a direct comparison between the two approaches. Equivalently, for each mass-ordering hypothesis we profile the 1D likelihood over $\Delta m^2_{32}$ by minimising the negative log-likelihood with respect to $\Delta m^2_{32}$, and we assign the ordering corresponding to the smaller profiled minimum.

A fully connected feed-forward NN is constructed, with two hidden layers, each containing 64 nodes and rectified linear unit (ReLU) activation functions. The benchmark parameter estimation study described in \Cref{subsec:benchmark} is a regression problem, for which the output layer uses a linear activation, where the model is trained using mean squared error loss. For the primary goal of MO estimation in \Cref{subsec:MO_estimate}, the problem is a binary classification, and the output layer employs a sigmoid activation to produce a probability score for the NO hypothesis. The model is trained using binary cross-entropy loss (maximum likelihood). Both models are optimised with the Adam algorithm~\cite{kingma2015adam}. All implementation is in Python using Keras~\cite{keras2015} and TensorFlow~\cite{tensorflow2015}, ensuring reproducibility and facilitating future pedagogical applications. 

The metrics compared for the MO determination study include the classification accuracy, the area under the ROC curve (AUC), and the $\mathrm{F_1}$ score. The $\mathrm{F_1}$ score is a metric for balancing precision (aka "purity" in physics parlance) and recall (aka "efficiency"), defined as $\mathrm{F_1}=\dfrac{2pr}{p+r}$, where $p$ is the precision and $r$ is the recall.

\begin{figure}[htpb!]
    \centering
    \includegraphics[width=\textwidth]{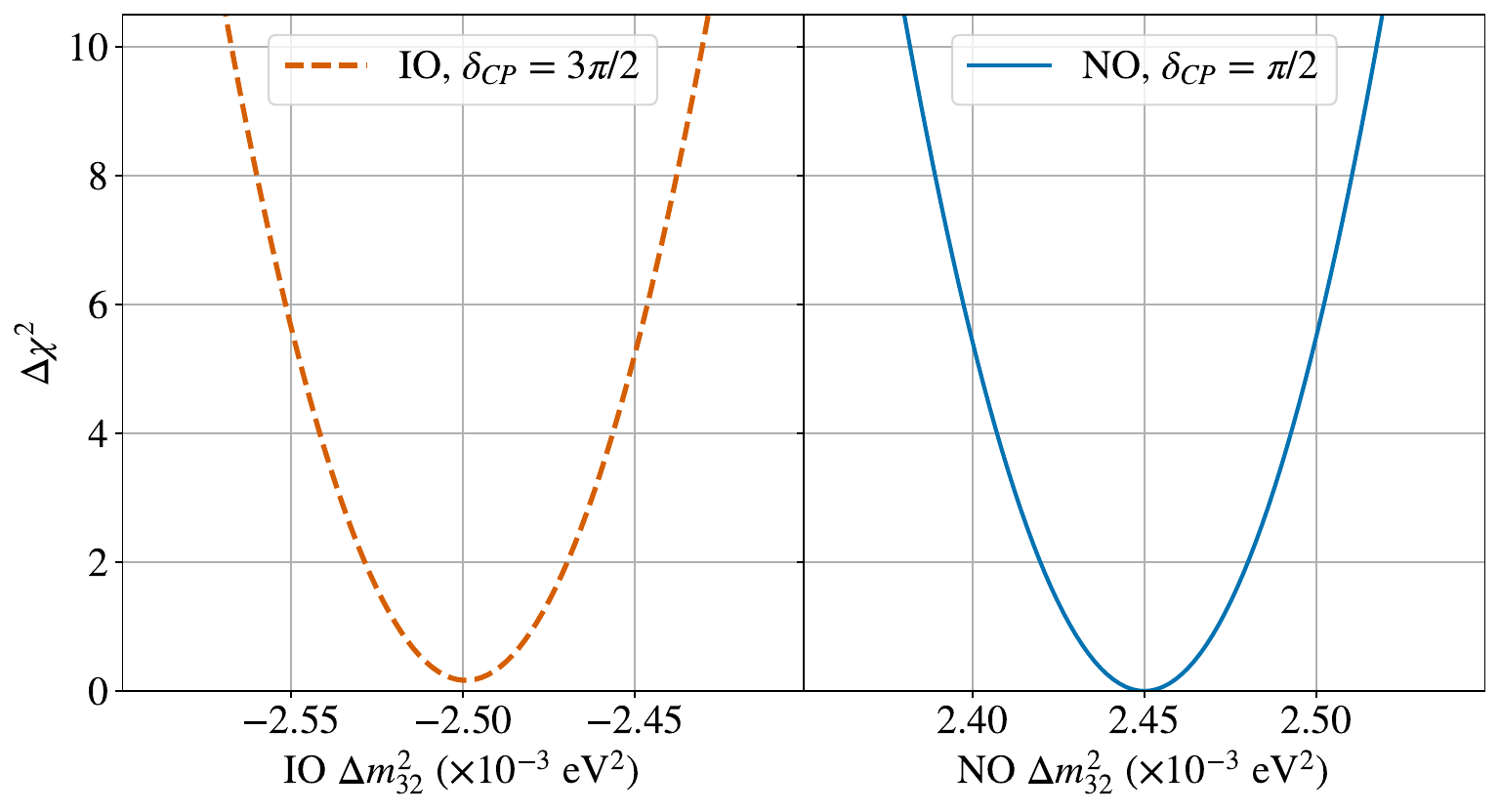}
    \caption{Difference in chi-squared, $\Delta\chi^{2}$, as a function of $\Delta m^{2}_{32}$ for both mass orderings at the ambiguity point ($\delta_{CP} = \pi/2$ for NO and $\delta_{CP} = 3\pi/2$ for IO). For each test hypothesis, $\Delta\chi^{2}$ is computed as the difference between the chi-squared value of the test histogram and that of a reference histogram generated under NO with $\Delta m^{2}_{32} = 2.45\times10^{-3}\,\text{eV}^2$. The minimum value under the IO assumption is $\Delta\chi^{2}_{\min} = 0.166$, which corresponds to an error rate of roughly 42\%~\cite{NMO_sens_quant:2014}, i.e., about 42\% of NO-like datasets would be misidentified as IO.}
    \label{fig:chi2}
\end{figure}

\Cref{fig:chi2} further illustrates the difficulty faced by long-baseline experiments such as NOvA when attempting to determine the neutrino mass ordering near ambiguity points. We compute a chi-squared value,
\chisq[test]\ for each sample array and mass ordering assumption by comparing them to a reference array \chisq[ref]\ generated with the NO assumption and with the PDG value of \dmatm. This is done separately for the NO and IO assumptions, and the difference \dchisq $\equiv$ \chisq[test] - \chisq[ref] is plotted as a function of \dmatm\ shown in \Cref{fig:chi2}. In practical terms, this means that about 42\% of datasets produced under NO would be misclassified as IO when using a standard histogram-based chi-squared comparison and assuming that either NO or IO is the correct model~\cite{NMO_sens_quant:2014}. The aim of this study is to explore whether a neural network can reduce this misclassification rate.

\section{Results}
\label{sec:results}

In this section, we present the performance of the NN trained on the synthetic datasets described in \Cref{sec:methods}. In \Cref{subsec:benchmark} we describe a benchmark study which compares the performance of the NN and LLH approaches in predicting the mass-squared splitting in a simplified `two-flavour' neutrino oscillation model. This simplified model is chosen for clarity, as the oscillation probability is simple enough to be written on a single line, \Cref{eq:2fprob}. Having successfully shown that a trained NN can match the performance of the LLH fit in determining the value of the two-flavour mass-squared splitting parameter, we proceed to the primary goal of this work: to assess the performance of an NN trained as an MO classifier, in \Cref{subsec:MO_estimate}. 

We report baseline results obtained with the full training dataset and default architecture, followed by an evaluation of classification metrics such as accuracy, AUC, and $\mathrm{F_1}$ score. This assessment aims to determine whether an NN fitter can match, or exceed, the neutrino mass-ordering sensitivity provided by a conventional LLH fitter.

\subsection{Benchmark Study: Estimation of $\Delta m^2$ in a two-flavour model.}
\label{subsec:benchmark}

To benchmark the NN against traditional statistical methods, we first compare its estimates of the mass-squared splitting with values obtained from LLH minimisation, using a simplified two-flavour survival probability in vacuum.

In this simplified model, the survival probability of a muon neutrino is described by:
\begin{equation}\label{eq:2fprob}
P(\nu_{\mu}\rightarrow\nu_{\mu})=1 - \sin^22\theta \sin^2(1.27 \Delta m^2 L/E), 
\end{equation}

where $\Delta m^2=2.45\times 10^{-3}$~eV$^2$ and $\sin^22\theta=0.9951$. For neutrinos of 2~GeV travelling 810~km, the survival probability is close to a local minimum. In this two flavour scenario, the MO choice is not relevant. 

\begin{figure}[htpb!]
    \centering
    \includegraphics[width=0.46\textwidth]{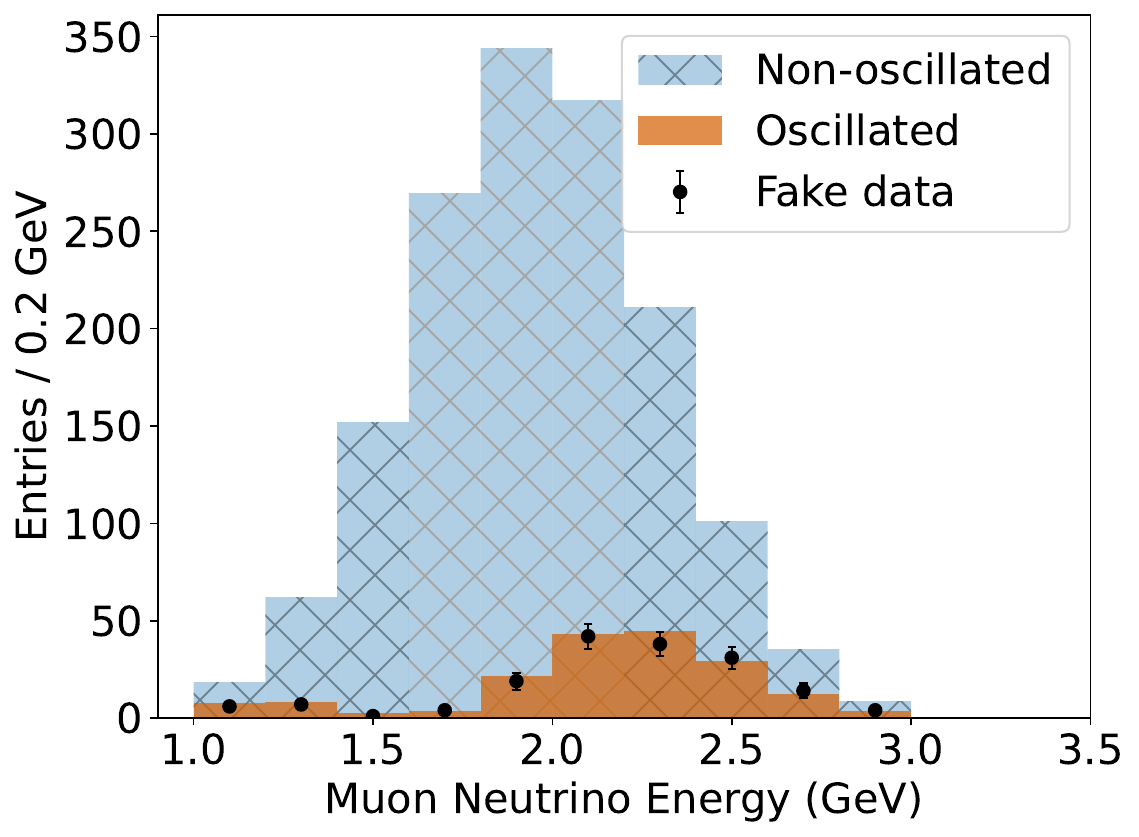}
    \includegraphics[width=0.44\textwidth]{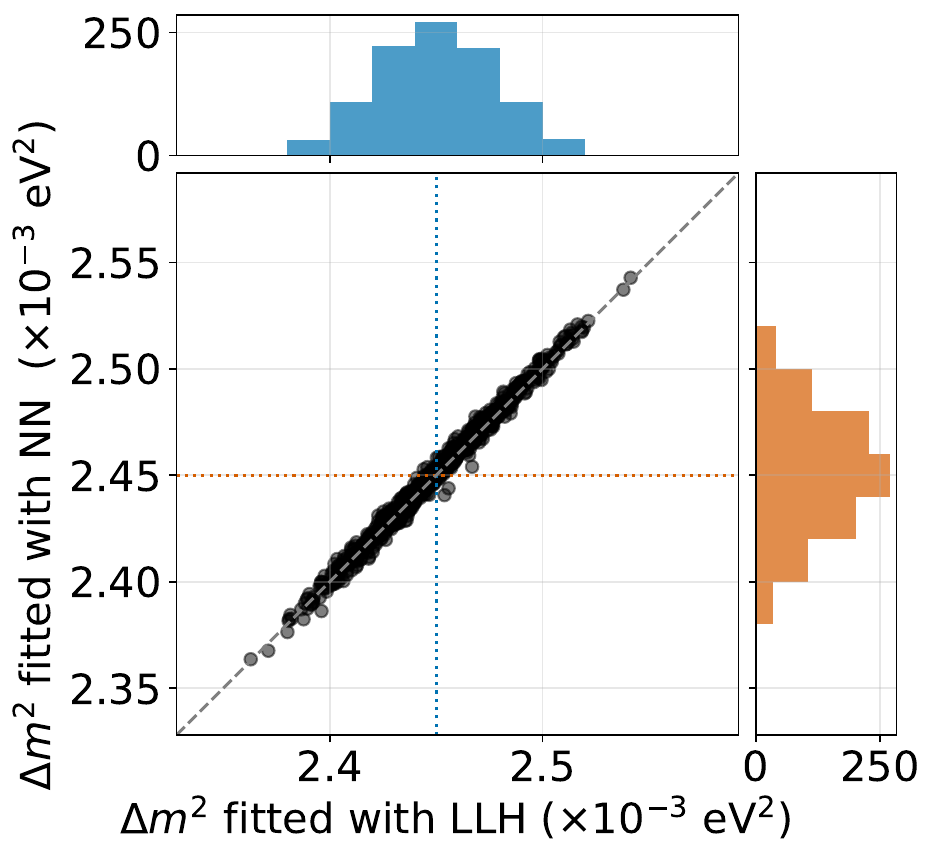}
    \caption{Left: The oscillated energy spectrum (orange) is obtained using the simplified two-flavour survival probability described in text, and assuming a baseline of 810~km. The black markers represent a pseudo-data energy spectrum, generated using Poisson fluctuations from the oscillated histogram. The blue hatched histogram shows the energy spectrum in the absence of oscillations. Right: The $\Delta m^2$ estimates from LLH (blue histogram) and NN (orange histogram) methods for the test sample of 1{,}000 pseudo-experiments. When both methods are applied to the same fluctuated neutrino energy spectrum, a strong correlation ($\rho=0.996$) is observed. The dotted blue and orange lines represent the true $\Delta m^2 = 2.45 \times 10^{-3}$~eV$^2$ used to generate the test sample histograms.
    }
    \label{fig:spec}
\end{figure}

The left panel of \Cref{fig:spec} shows representative spectra with and without oscillations. The pseudo-data histogram is a proxy for the measurement of an oscillation experiment. The NN training sample comprises one million measurements in this histogram format, generated for $\Delta m^2$ values in the range $[1.0,\,4.0]\times 10^{-3}$~eV$^2$. 

The sample generated to test the trained NN comprises one thousand measurements with fixed  $\Delta m^2 = 2.45 \times 10^{-3}$~eV$^2$. Both the training and test samples include bin-by-bin Poisson fluctuations. 

The trained NN extracts a value of $\Delta m^2 = (2.451 \pm 0.028) \times 10^{-3}$~eV$^2$ from the test sample, where the central value is the mean and the uncertainty is the standard deviation of the extracted values. The LLH fit yields $(2.450 \pm 0.028) \times 10^{-3}$~eV$^2$, identical to the NN prediction at this precision. 

The histograms of predicted $\Delta m^2$ values are shown in the right panel of \Cref{fig:spec}, along with their correlation. The Pearson correlation coefficient between the LLH and NN outputs is 0.996.

Neutrino oscillation experiments make a single measurement of a given parameter, rather than the set of 1,000 measurements in the test sample used here. The LLH fit provides both a central value and an uncertainty for a single measurement, but the trained NN cannot provide an uncertainty from a single value. To mimic the conditions of a single experimental realisation while also providing an NN uncertainty on the parameter estimate, a bootstrap sample is used. A single histogram is selected at random from the test sample, and is used to generate a bootstrap sample of 1,000 measurements. The bootstrap distribution of NN-extracted $\Delta m^2$ values has a median $\Delta m^2 = 2.402 \times 10^{-3}$~eV$^2$ and a 1$\sigma$ interval of $\pm 0.031 \times 10^{-3}$~eV$^2$. A LLH fit to this sample yields a consistent median and 1$\sigma$ interval. The LLH fit does provide an uncertainty for the single experiment, $(2.406 \pm 0.026) \times 10^{-3}$~eV$^2$, consistent with the expectation. 

Taken together, the tests above demonstrate that the NN-based approach can infer the oscillation parameters with a precision comparable to that of standard likelihood fits.

\subsection{Estimation of the Neutrino Mass Ordering}
\label{subsec:MO_estimate}
 The synthetic data used for this study are those described in \Cref{subsec:data_gen}; four histograms, each containing the energy bin counts of $\nu_\mu$, $\bar{\nu}_\mu$, $\nu_e$, and $\bar{\nu}_e$, with full three-flavour oscillation probabilities and Poisson fluctuations applied. The NN is trained using two million measurements, with $\Delta m^2_{32}$ randomly varied between $[-2.65,\,-2.41]\times 10^{-3}$~eV$^2$ and $[2.33,\,2.57]\times 10^{-3}$~eV$^2$, covering a larger range than in \Cref{fig:chi2}. The value of $\delta_{\rm CP}$ is set to $\pi/2$ ($3\pi/2$) for normal (inverted) mass ordering so that the ambiguous region described in \Cref{sec:methods} is covered. This trained NN returns a value between zero and one: the closer the output is to one (zero), the higher the confidence that the measurement is compatible with normal (inverted) neutrino mass ordering. 

\begin{figure}[htpb!]
    \centering
    \includegraphics[width=0.75\textwidth]{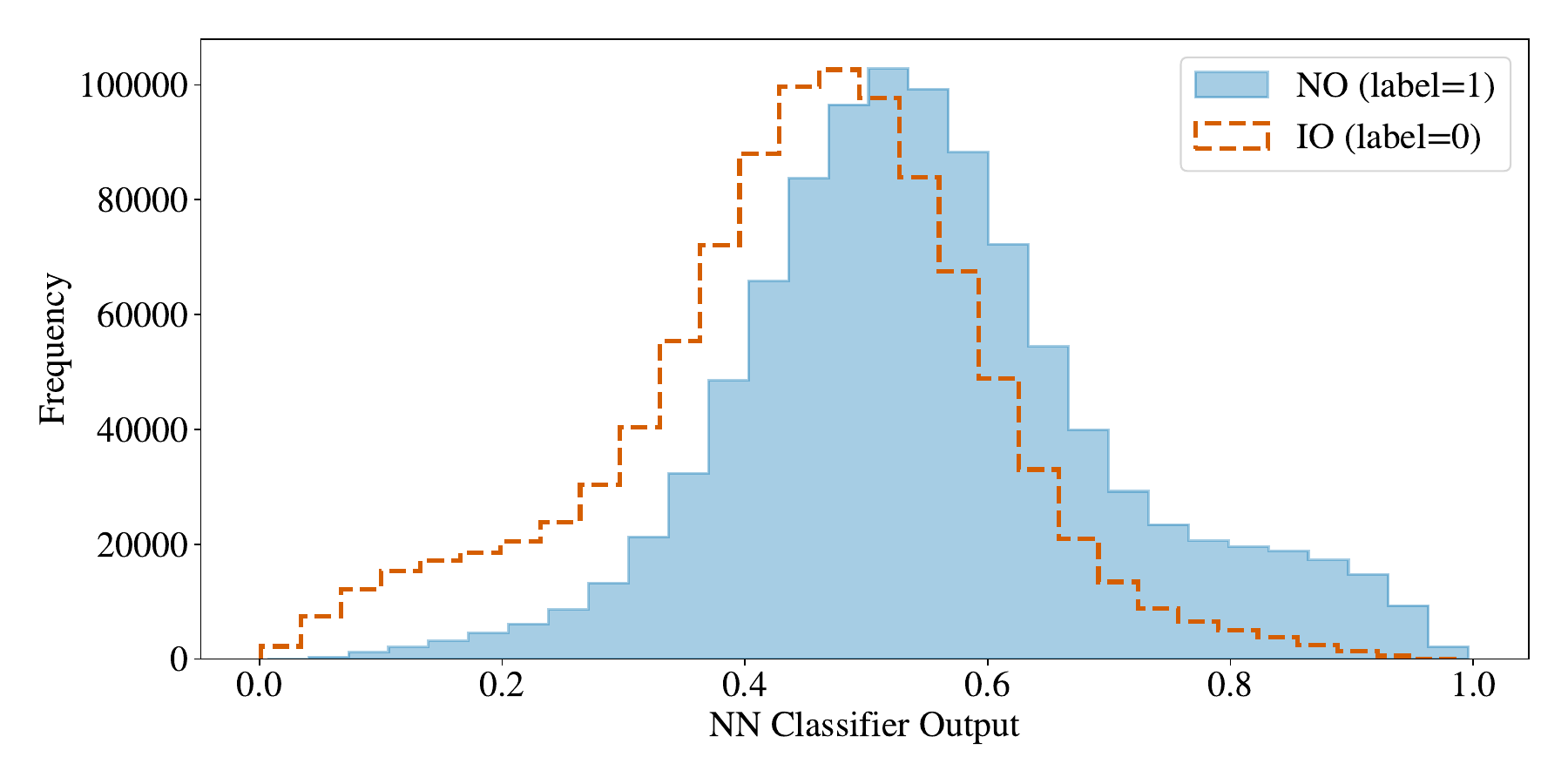}
    \includegraphics[width=0.75\textwidth]{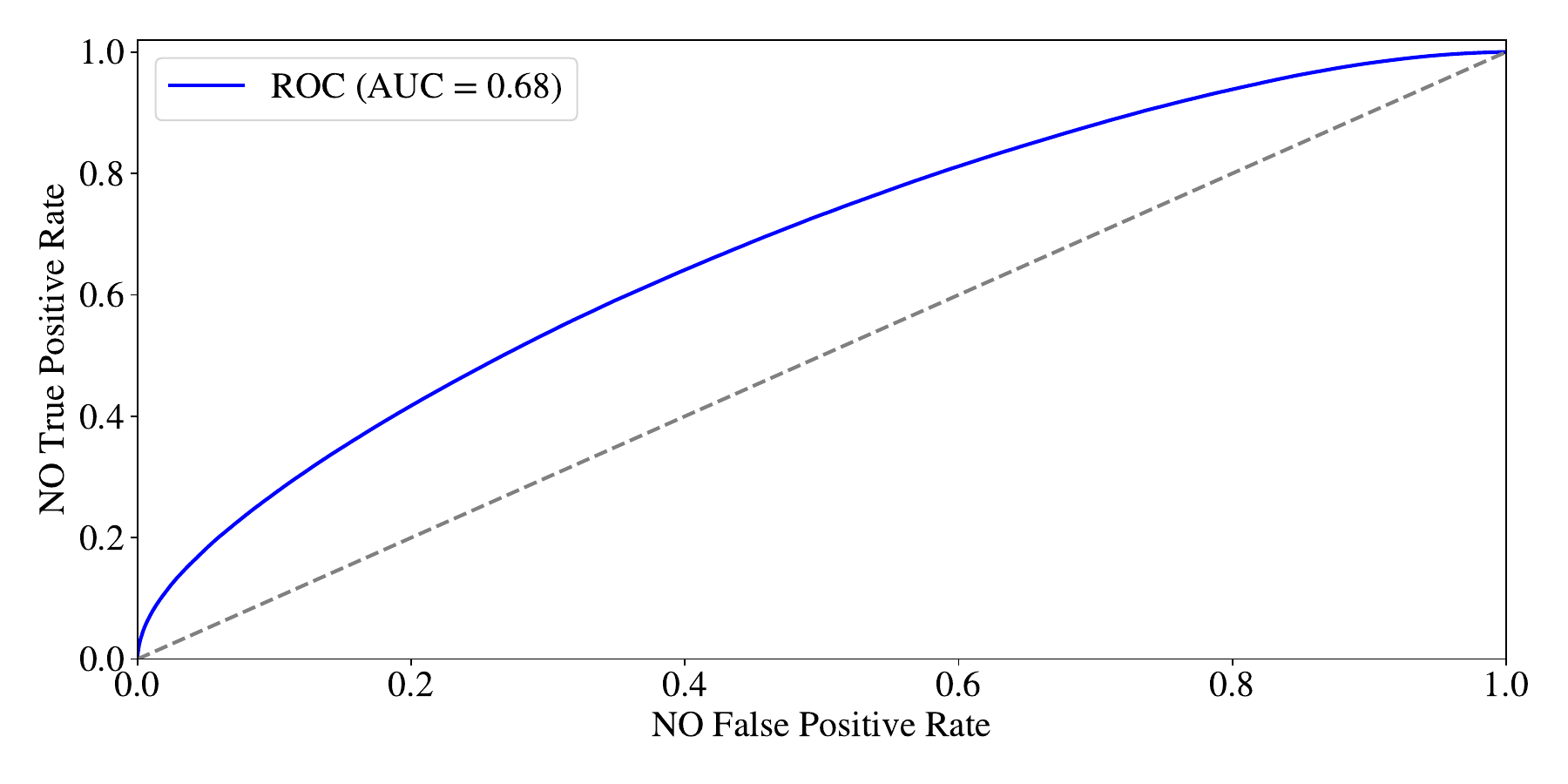}
    \caption{Results of NN-based MO classification using the training sample. Top: output probability of the NN classifier for NO and IO hypotheses. Blue (orange dashed-outline) histograms correspond to measurements generated with true NO (IO). Bottom: ROC curve showing true positive rate (NO classified as NO) versus false positive rate (IO classified as NO) across thresholds of the NN classifier output.}
    \label{fig:model_output}
\end{figure}

\Cref{fig:model_output} presents the distribution of the output of the NN classifier (top) and the corresponding ROC curve (bottom). The figure indicates that the trained NN shows some ability to distinguish between normal and inverted neutrino mass ordering, given the parameters and bins chosen for this study. For this baseline configuration, the area under the ROC curve (AUC) is 0.68, indicating better-than-random discrimination power. An AUC of 1 corresponds to a perfect classifier, while an AUC of 0.5 reflects performance equivalent to random guessing. Values below 0.5 indicate that the classifier tends to systematically invert the labelling, meaning that its predictions would provide useful information only after flipping the output.

\begin{figure}[htpb!]
    \centering
    \includegraphics[width=0.75\textwidth]{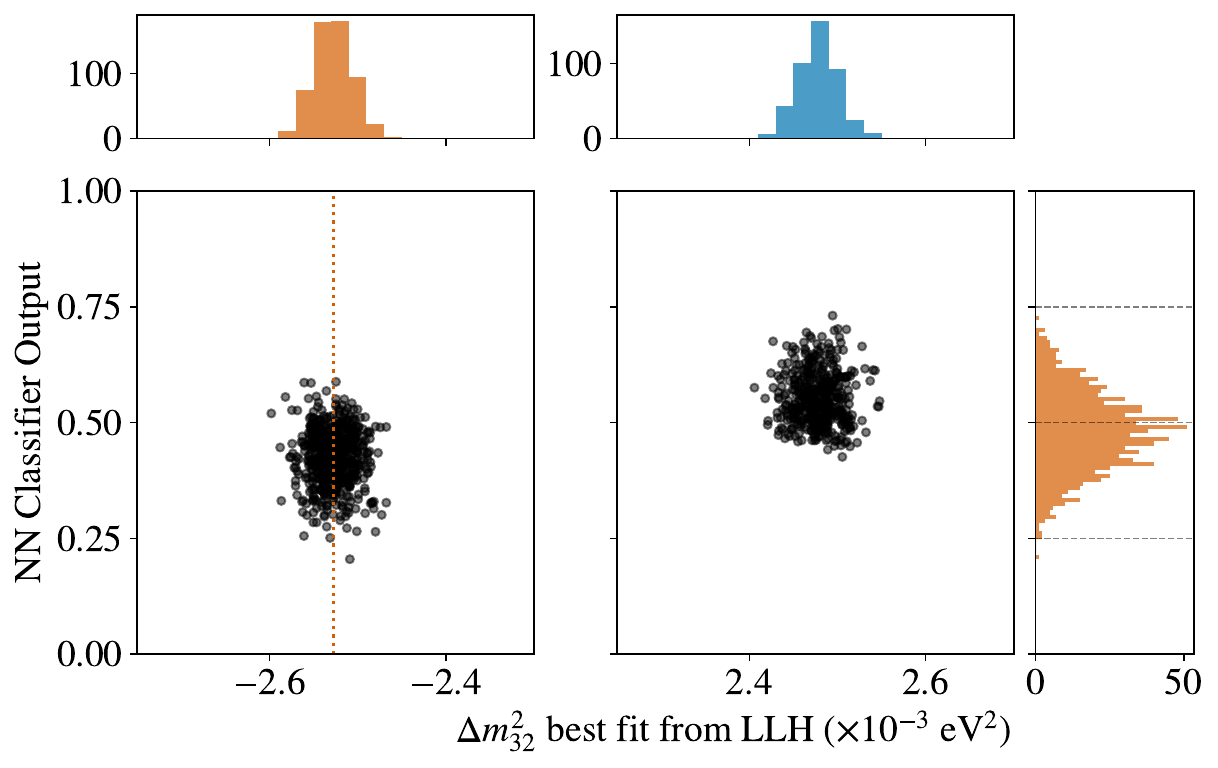}
    \caption{Comparison between the NN model and standard LLH minimisation using an independent test sample of 1{,}000 Poisson-fluctuated synthetic datasets with fixed $\Delta m^2_{32}= -2.527\times10^{-3}$~eV$^{2}$ (IO). The x-axis shows the $\Delta m^2_{32}$ values from the LLH minimisation, where positive (negative) values are for normal (inverted) ordering. The y-axis is for the output of the NN model, with values closer to 0 (1) indicating a preference for IO (NO). The projections at the top show the LLH histograms. The vertical dotted orange line indicates the true central value of $\Delta m^2_{32}$. The average of the NN output has a slight preference for the IO scenario, indicated by values closer to zero.}
    \label{fig:NN_llh_test}
\end{figure}

The performance is further examined using an independent test sample (i.e., not used during NN training) of 1,000 NO-true plus 1,000 IO-true measurements (2,000 in total), with the $\Delta m^2_{32}$ values fixed to the relevant PDG values given in \Cref{tab:params}, and Poisson fluctuations applied. \Cref{fig:NN_llh_test} compares the NN output and LLH minimisation for the true IO test sample. The NN output shows a slight preference for the IO scenario, indicated by values closer to zero (average lower than 0.5). Similar behaviour is seen in the LLH minimisation, where negative values of $\Delta m^2_{32}$ are more probable.

\begin{figure}[htpb!]
    \centering
    \includegraphics[width=0.75\textwidth]{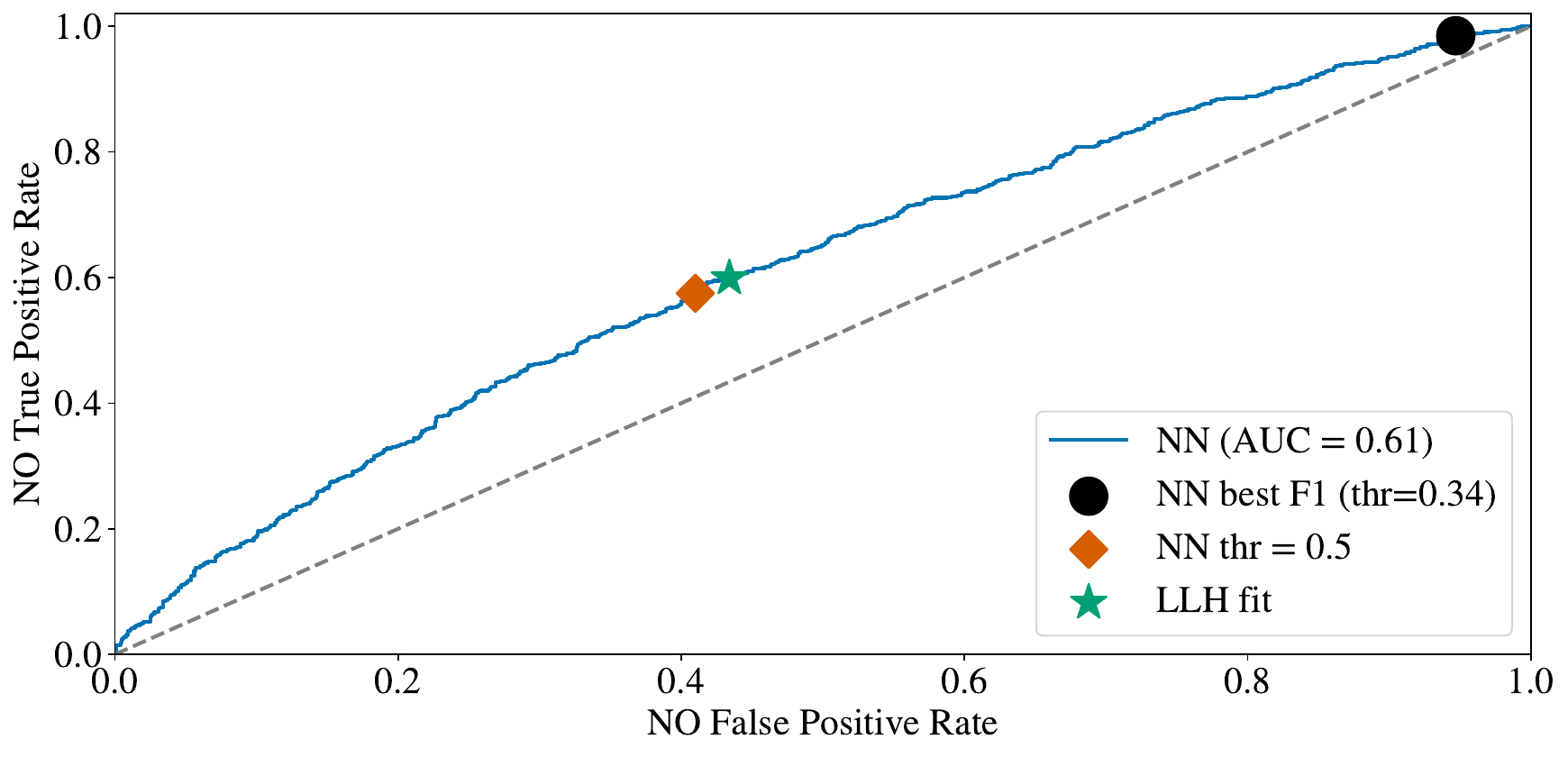}
    \caption{ROC curve of the NN model applied to a test sample and standard LLH minimisation point. The data used to generate this curve consist of 1{,}000 NO samples and 1{,}000 IO samples, with fixed parameters as described in the text. The black filled marker shows the operating point given by the NN-classifier threshold that maximises the F1 score. The orange diamond marker shows the NN at the simple symmetric threshold of 0.5. The bluish-green star marks the operating point of the standard LLH minimisation.
    }
    \label{fig:roc_test}
\end{figure}

In this validation sample of NO and IO measurements, the NN achieves an AUC of 0.61, with the maximal $\mathrm{F_1}$ score obtained at a classifier output threshold of 0.34 (i.e., samples with an NN output $>0.34$ are classified as NO). The true positive rate (NO classified as NO) of this validation sample is 98.5\% while the false positive rate (IO classified as NO) is 94.7\%. Here the F$_1$-optimising threshold favours high recall for NO at the expense of precision, reflecting the substantial overlap between the NN score distributions for the fixed-parameter test sample. \Cref{fig:roc_test} shows the ROC curve and the false/true positive rates for this test sample. The figure also shows the results obtained with the LLH minimisation applied to the same samples. The LLH method achieves a true positive rate of 59.9\% and a false positive rate of 43.4\%. This corresponds to a single point with which to compare the NN ROC curve. 
For comparison, \Cref{fig:roc_test} also marks the NN operating point at the na\"{\i}ve symmetric threshold of 0.5 (orange diamond), where its true and false positive rates (57.5\% and 41.0\% respectively) closely match the LLH operating point. 

This study does not find evidence that the NN exhibits significantly stronger classification power than the LLH fit. However, unlike the LLH approach, the NN provides a continuous output that enables ROC-curve construction and permits the selection of operating points with higher true positive rates where appropriate. 

\section{Discussion}
\label{sec:concl}

The neutrino mass ordering remains one of the most important open questions in neutrino physics, with implications extending beyond particle physics to astrophysics and cosmology. The next generation of neutrino experiments is expected to provide a definitive answer within the coming decade. In parallel, there is substantial interest in exploring new methods to extract maximal information from existing data.

In this work, we investigated whether machine-learning techniques, specifically a feed-forward neural network, could improve sensitivity to the mass ordering relative to traditional approaches such as log-likelihood minimisation. Using synthetic datasets based on the NOvA experimental configuration, we trained and evaluated the model under realistic conditions that incorporate matter effects and statistical fluctuations.

The results of this toy study show that a neural network can match the classification performance of a log-likelihood fit. This outcome suggests that oscillated (anti)neutrino energy spectra do not contain sizeable hidden correlations beyond those exploited by conventional techniques. Nevertheless, the NN approach offers practical advantages, including flexibility in threshold selection through ROC curves, fast results after a model is trained, and ease of integration into modern data-analysis workflows.

Beyond its immediate findings, this study provides a valuable consistency check and demonstrates the feasibility of using machine learning for fitting in neutrino oscillation analyses. The framework developed here can be extended to more complex scenarios, such as joint optimisation of oscillation parameters or inclusion of systematic uncertainties.

Several refinements to this toy study could be explored in future work. The Gaussian flux approximation and the 22-bin layout used here were chosen primarily for simplicity. The use of a more accurate flux model (the NOvA beam flux is narrow-band) and a finer or jointly-learned binning, tuned for classifier performance, could reveal additional spectral structure for the network to exploit.  Alternative architectures, such as convolutional or graph-based networks, may extract subtle features from high-dimensional oscillation data more effectively than the feed-forward network used here.

\section*{Acknowledgments}
The authors thank Prof. Mark Messier for suggesting the analysis presented in this work and insightful discussions on the methodology, and the NOvA Collaboration for their valuable feedback throughout the development of this study. This work was supported by the Science and Technology Facilities Council (STFC) of UK Research and Innovation (UKRI) under the grant ST/W000512/1.

\section*{Code and data availability}
  The analysis code that reproduces all results presented in this manuscript is publicly available at \url{https://github.com/universityofsussex-mps/nn-vs-llh-mass-ordering}.  The trained neural-network
  models and the synthetic datasets generated and analysed for this study are archived on Zenodo~\cite{data_zenodo26}.

\bibliographystyle{unsrt}
\bibliography{references.bib}

\end{document}